\def\prg#1{\medskip{\bf #1}}
\def\lra{\leftrightarrow}        
               \def\pd{\partial}
\def\cs{{\scriptstyle\rm CS}}    \def\mb#1{{\boldmath$#1$}}
\def\dis{\displaystyle}          \def\dfrac{\dis\frac}

       \def\tI{{\tilde I}}
\def\tR{{\tilde R}}
\def\hB{{\hat B}}     \def\hOm{{\hat\Omega}}
\def\bA{{\bar A}}     \def\bu{{\bar u}}
\def\cA{{\cal A}}     \def\bcA{{\bar{\cal A}}}
     
\def\bG{{\bar G}}     \def\bg{{\bar g}}
\def\bK{{\bar K}}     
\def\cL{{\cal L}}     \def\cM{{\cal M }}
\def\cO{{\cal O}}     \def\cE{{\cal E}}

\def\m{\mu}           \def\n{\nu}           \def\k{\kappa}
        \def\g{\gamma}        \def\d{\delta}
\def\S{\Sigma}        \def\s{\sigma}        \def\t{\tau}
\def\a{\alpha}        \def\b{\beta}         \def\th{\theta}
\def\vphi{\varphi}    \def\ve{\varepsilon}  \def\p{\pi}
\def\r{\rho}          \def\D{\Delta}        
\def\l{\lambda}       \def\om{\omega}       \def\Om{\Omega}
  \def\Th{\theta}

\def\hs{\hspace{1.3pt}}               \def\Tr{{\rm\hs Tr\hs}}
\def\nn{\nonumber}
\def\be{\begin{equation}}             \def\ee{\end{equation}}
\def\ba#1{\begin{array}{#1}}          \def\ea{\end{array}}
\def\bea{\begin{eqnarray} }           \def\eea{\end{eqnarray} }
\def\beann{\begin{eqnarray*} }        \def\eeann{\end{eqnarray*} }
\def\beal{\begin{eqalign}}            \def\eeal{\end{eqalign}}
\def\lab#1{\label{eq:#1}}             \def\eq#1{(\ref{eq:#1})}
\def\bsubeq{\begin{mathletters}}      \def\esubeq{\end{mathletters}}
\def\bitem{\begin{itemize}}           \def\eitem{\end{itemize}}

\documentstyle[aps,preprint,eqsecnum]{revtex}    

\begin{document}
\tighten

\title{Asymptotic dynamics in 3D gravity with torsion}

\author{M.\ Blagojevi\'c and M. Vasili\'c\thanks{Email
        addresses:  mb@phy.bg.ac.yu, mvasilic@phy.bg.ac.yu}}
\address{Institute of Physics, P.O.Box 57, 11001 Belgrade, Serbia}

\maketitle

\begin{abstract}
We study the nature of boundary dynamics in the teleparallel 3D
gravity. The asymptotic field equations with anti-de Sitter boundary
conditions yield only two non-trivial boundary modes, related to
a conformal field theory with classical central charge. After showing
that the teleparallel gravity can be formulated as a Chern-Simons
theory, we identify dynamical structure at the boundary as the
Liouville theory.
\end{abstract}


\section{Introduction}

The growth of interest in three-dimensional (3D) gravity in the past
twenty years is motivated by the hope that such a simple model might
help us to get a better understanding of the intricate structure of the
realistic four-dimensional theory \cite{1,2}. Particularly important
advances along these lines were achieved by clarifying the asymptotic
structure of 3D gravity \cite{3}, and formulating 3D gravity as a
Chern-Simons theory \cite{4}. These results have a significant
influence on our understanding of the quantum nature of 3D black holes
\cite{5,6,7,8,9,10,11,12}. However, the analysis of these issues has
been carried out almost exclusively in the realm of {\it Riemannian\/}
geometry of general relativity (GR). The investigation of the
asymptotic structure of 3D gravity in the context of {\it
Riemann-Cartan\/} geometry, a geometry possessing both the curvature
and the torsion \cite{13,14}, has been initiated in our resent
publication \cite{15}.

The asymptotic content of gravity is most clearly seen in
topological theories, where the non-trivial dynamics can exist only
at the boundary. Mielke and Baekler \cite{16,17} proposed a general
action for topological Riemann-Cartan 3D gravity. The analysis of
its classical solutions has been recently completed in Ref.
\cite{18}. For a specific choice of free parameters, this theory
leads to the {\it teleparallel\/} (or Weitzenb\"ock) geometry of
spacetime \cite{19,20,21,14}, in which curvature vanishes and
torsion remains non-trivial. This limit is particularly simple
framework for studying the influence of torsion on the boundary
dynamics. After having clarified the nature of asymptotic symmetry
in the teleparallel 3D gravity \cite{15}, we now continue with the
analysis of the related asymptotic dynamics.

Riemann-Cartan theory can be formulated as Poincar\'e gauge theory
\cite{13,14}. Basic gravitational variables are the triad field
$b^i{_\m}$ and the Lorentz connection $A^{ij}{}_\m$, and the related
field strengths are the torsion $T^i{}_{\m\n}$ and the curvature
$R^{ij}{}_{\m\n}$. In 3D, we can simplify the notation by introducing
$$
\om_{i\m}=-\frac{1}{2}\ve_{ijk}A^{jk}{}_\m\, ,\qquad
R_{i\m\n}=-\frac{1}{2}\ve_{ijk}R^{jk}{}_{\m\n}\, .
$$
Then, the gauge transformations of the fields, local Lorentz
transformations and local translations  with parameters $\th^i$ and
$\xi^\m$, take the form
\bsubeq\lab{1.1}
\bea
\d_0 b^i{_\m}&=& \ve^{ijk}\th_j b_{k\m}-\xi^{\r}{}_{,\,\m}b^i{_\r}
               -\xi^{\r}\pd_\r b^i{}_{\m}                   \nn\\
\d_0\om^i{_\m}&=& -\nabla_\m\th^i-\xi^{\r}{}_{,\,\m}\om^i{_\r}
                  -\xi^{\r}\pd_\r\om^i{}_{\m}\, ,        \lab{1.1a}
\eea
where $\nabla_\m\th^i=\pd_\m\th^i+\ve^i{}_{mn}\om^m{_\m}\th^n$, and the
field strengths are given as
\bea
&&T^i{}_{\m\n}=\pd_\m b^i{_\n}+\ve^{ijk}\om_{j\m}b_{k\n}
               -(\m\lra\n)\, ,                    \nn\\
&&R^i{}_{\m\n}=\pd_\m\om^i{_\n}-\pd_\n\om^i{_\m}
               +\ve^{ijk}\om_{j\m}\om_{k\n}\, .           \lab{1.1b}
\eea
\esubeq
Starting from the general Mielke-Baekler action \cite{16,17}, the
gravitational sector of our Riemann-Cartan theory is assumed to have a
specific form, suitable for studying the role of torsion at the
boundary \cite{15}:
\bea
&&I_G=\int_\cM d^3\xi\,\cL_G\, ,    \nn\\
&&\cL_G=a\left[\ve^{\r\m\n} b^i{_\r}\left(R_{i\m\n}
         -\frac{1}{\ell}T_{i\m\n}\right)
         +\frac{8}{\ell^2}\,b\right]\, .                   \lab{1.2}
\eea
Here, $a=1/16\pi G$, $\ell$ is a constant, $b=\det(b^i{_\m})$, and
$\cM$ is a manifold with the topology $R\times\S$, where $R$ is
interpreted as time, and $\S$ is a spatial manifold with the asymptotic
region of $R^2$ type. The related field equations in vacuum,
\be
R^i{}_{\m\n}=0\, ,\qquad
T^i{}_{\m\n}=\frac{2}{\ell}\,\ve^i{}_{jk}b^j{_\m}b^k{_\n}\,,\lab{1.3}
\ee
imply that the geometry of spacetime at large distances, outside the
regions occupied by matter, is teleparallel.

In addition to the field equations, asymptotic conditions play an
equally important role in defining the dynamical structure of a field
theory. Our study of the {\it asymptotic symmetry\/} of the
teleparallel theory \eq{1.2} is based on the concept of an
asymptotically anti-de Sitter (AdS) configuration of fields, which is
precisely defined by the requirements (4.2) in Ref. \cite{15}. The
canonical realization of the asymptotic symmetry is found to be the
same as in GR --- the conformal symmetry with classical central charge
$c=3\ell/2G$. The subject of the present paper is to precisely identify
the related {\it asymptotic dynamics\/}, whereupon we shall be able to
completely understand the dynamical role of torsion at the boundary of
the teleparallel spacetime.

The paper is organized as follows. In Sec. II we study the dynamical
content of the gravitational field equations in the asymptotic
(boundary) region, and show that it can be described by two independent
boundary modes --- a left moving and a right moving mode. In Sec. II we
find a simple change of variables,
$(b^i{_\m},\om^i{_\m})\to(A^i{_\m},\bA^i{_\m})$, which transforms the
teleparallel theory \eq{1.2} into a Chern-Simons theory, in analogy
with Witten's result for GR \cite{4}. In Sec. IV, using the new
variables we establish the equivalence of the teleparallel boundary
dynamics with Liouville theory, again in analogy with GR \cite{5}.
These results lead us to an important conclusion: in spite of different
geometric content, the boundary dynamics in the teleparallel theory is
exactly the same as in GR. Finally,  Sec. V is devoted to concluding
remarks, and appendices contain some technical details.

Our conventions are the same as in Ref. \cite{15}: the Latin indices
$(i,j,k,...)$ refer to the local Lorentz frame, the Greek indices
$(\m,\n,\r,...)$ refer to the coordinate frame, and both run over
$0,1,2$; $\eta_{ij}=(+,-,-)$ and $g_{\m\n}=b^i{_\m}b^j{_\n}\eta_{ij}$
are metric components in the local Lorentz and coordinate frame,
respectively; totally antisymmetric tensor $\ve^{ijk}$ and the related
tensor density $\ve^{\m\n\r}$ are both normalized by $\ve^{012}=+1$.

\section{Boundary dynamics 1}

In Riemann-Cartan spacetime, we can use a well known geometric
identity to express the curvature tensor in terms of its
Riemannian piece $\tR_{\m\n\l\r}$ and the contortion \cite{14,15}.
Then, the field equations \eq{1.3} can be rewritten as
$$
\tR_{\m\n\l\r}=\frac{1}{\ell^2}\left(g_{\m\l}g_{\n\r}
               -g_{\m\r}g_{\n\l}\right)\, , \qquad
T^i{}_{\m\n}=\frac{2}{\ell}\,\ve^i{}_{jk}b^j{_\m}b^k{_\n}\, .
$$
The torsion equation eliminates the connection $\om^i{_\m}$ as an
independent dynamical variable, and we are left with the maximally
symmetric metric $g_{\m\n}$, corresponding to the AdS space
\cite{22}. As maximally symmetric spaces are locally unique, we
conclude that our theory carries no local degrees of freedom.
Instead, all the solutions of the theory are gauge equivalent, at
least in local regions of spacetime.

Although the total number of local degrees of freedom is zero, the
theory turns out to have a rich asymptotic structure. Namely, after
adopting a suitable set of asymptotic conditions (defined by the AdS
asymptotic configurations of fields), one finds that the complete gauge
symmetry of the theory can be naturally split up into two parts: a) the
{\it asymptotic symmetry\/} and b) the remaining {\it proper gauge
symmetry\/} \cite{15,23,3}. The first acts on the boundary, has the
structure isomorphic to the two-dimensional conformal group, and is
produced by the generators with non-vanishing conserved charges. The
second acts on the spacetime interior, and has generators with
vanishing conserved charges.

To a great extent, the physical (geometric) interpretation of the above
symmetry structure is motivated by the related quantum theory of
gravity \cite{2,5,6,7,8,9,10,11}. At the quantum level, the asymptotic
symmetry is broken (conformal anomaly), and therefore, it should not be
treated as a genuine gauge symmetry, but as a physical one. At the same
time, the proper gauge symmetry remains truly non-physical. The
existence of the proper gauge freedom implies that the whole interior
of spacetime is pure gauge, and carries no local dynamical degrees of
freedom. These are located only on the boundary of the underlying
spacetime. Classical theory with such an interpretation of its symmetry
structure has a direct relevance for the corresponding quantum theory
(see also Refs. \cite{24}).

The subject of the present paper is to identify the boundary degrees of
freedom associated with the breakdown of gauge invariance at the
boundary. Their number is expected to be the same as the number of
parameters of the conformal symmetry. Aware of the fact that the
spacetime interior carries no local degrees of freedom, we shall focus
our attention solely on the dynamical properties at the boundary.

The asymptotic configuration of the basic dynamical variables is
defined using the concept of the AdS asymptotic behavior \cite{3,23}.
It is given by the expressions \cite{15}
\bsubeq\lab{2.1}
\be
b^i{_\m}= \left( \ba{ccc}
       \dfrac{r}{\ell}+\dfrac{\ell}{r}\,\hB^0{_0} &
       \dfrac{\ell^4}{r^4}\,\hB^0{_1} &
       \dfrac{\ell}{r}\,\hB^0{_2}             \\ [2.2ex]
       \dfrac{\ell^2}{r^2}\,\hB^1{_0} &
       \dfrac{\ell}{r}+\dfrac{\ell^3}{r^3}\,\hB^1{_1} &
       \dfrac{\ell^2}{r^2}\,\hB^1{_2}         \\ [2.2ex]
       \dfrac{\ell}{r}\,\hat B^2{_0} &
       \dfrac{\ell^4}{r^4}\,\hB^2{_1} &
       r+\dfrac{\ell}{r}\,\hB^2{_2}           \\[1ex]
             \ea \right)  \, ,                           \lab{2.1a}
\ee
\be
\om^i{_\m}=\left( \ba{ccc}
       \dfrac{r}{\ell^2}+\dfrac{\ell}{r}\,\hOm^0{_0} &
       \dfrac{\ell^4}{r^4}\,\hOm^0{_1} &
       -\dfrac{r}{\ell}+\dfrac{\ell}{r}\,\hat \Om^0{_2} \\[2.2ex]
       \dfrac{\ell^2}{r^2}\,\hOm^1{_0} &
       \dfrac{1}{r}+\dfrac{\ell^3}{r^3}\,\hat \Om^1{_1} &
       \dfrac{\ell^2}{r^2}\,\hOm^1{_2}                 \\[2.2ex]
       -\dfrac{r}{\ell^2}+\dfrac{\ell}{r}\,\hOm^2{_0} &
       \dfrac{\ell^4}{r^4}\,\hOm^2{_1} &
       \dfrac{r}{\ell}+\dfrac{\ell}{r}\,\hOm^2{_2}     \\[1ex]
             \ea \right)  \, ,                            \lab{2.1b}
\ee
\esubeq
where $x^\m=(t,r,\vphi)$.  In Ref. \cite{15}, the variables $\om^0{_1}$
and $\om^2{_1}$ behaved as $1/r^2$ at spatial infinity. However,
solving the constraints of the theory in the asymptotic region, one
discovers $\om^0{_1},\,\om^2{_1}=\cO_4$ ($\cO_n$ denotes a quantity
that tends to zero as $1/r^n$ or faster when $r\to\infty$). We decided
to take this result into account from the very beginning. Particularly
important solution of the type \eq{2.1} is the black hole solution
\cite{25}.

The new variables $\hB^i{_\m}$ and $\hOm^i{_\m}$ are defined to be of
the $\cO_0$ type. In what follows, we shall be interested only in the
lowest order terms (zero modes) of $\hB$, $\hOm$, whereas the higher
order terms are expected to be pure gauge terms, hence physically
irrelevant. As we shall see, the forthcoming results will justify this
approach.

The adopted boundary conditions \eq{2.1} restrict the gauge parameters
$\xi^\m,\th^i$ to have the form \cite{15}
\be
\ba{ll}
\xi^0=\ell\,T+\dfrac{\ell^5}{2r^2}\,T_{,00}
      +\dfrac{\ell^4}{r^4}\,\hat\xi^0 \,,\quad       &
\th^0=-\dfrac{\ell^2}{r}\,T_{,02}
      +\dfrac{\ell^3}{r^3}\,\hat\th^0\,,                \\[1.7ex]
\xi^1=-r\ell\,T_{,0}+\dfrac{\ell}{r}\,\hat\xi^1 \,,   &
\th^1=T_{,2}+\dfrac{\ell^2}{r^2}\,\hat\th^1\,,          \\[1.7ex]
\xi^2=S-\dfrac{\ell^2}{2r^2}\,S_{,22}
           +\dfrac{\ell^4}{r^4}\,\hat\xi^2\,,         &
\th^2=\dfrac{\ell^3}{r}\,T_{,00}
      +\dfrac{\ell^3}{r^3}\,\hat\th^2                   \\[1ex]
\ea                                                       \lab{2.2}
\ee
The parameters $T$ and $S$ satisfy the equations $\pd_{\pm}(T\mp S)=0$
and define the asymptotic symmetry group of the theory (the conformal
group with classical central charge $3\ell/2G$). The proper gauge
transformations are defined by the remaining parameters $\hat\xi^\m$,
$\hat\th^i$, which are arbitrary functions of the $\cO_0$ type. In what
follows, the proper gauge parameters $\hat\xi^\m$, $\hat\th^i$ will be
used to remove the unphysical modes at the boundary.

To simplify the notation, we utilize the light-cone basis to define a)
the coordinate and b) local Lorentz light-cone components of vectors:
\bsubeq
\bea
V^\m:&&\quad V^\pm=\frac{1}{\ell}V^0\pm V^2\, , \quad\,
      V_\pm=\frac{1}{2}\left(\ell V_0\pm V_2\right)\, ,\lab{2.3a}\\
U^i:&&\quad U^\pm=U^0\pm U^2\, , \qquad
      U_\pm=\frac{1}{2}\left(U_0\pm U_2\right)\, .     \lab{2.3b}
\eea
\esubeq
Note that with these conventions, the light-cone coordinates are
dimensionless: $x^\pm=t/\ell\pm\vphi$.

The action of the proper gauge group on $\hB$ and $\hOm$ can be
found from the general transformation law \eq{1.1a}. It is given
in Appendix A, and can be used to fix the proper gauge freedom in
the lowest order. In particular, we can choose the following six
gauge conditions:
\bea
&&\hB^{\pm}_1 = \cO_1 \,,\quad  \hB^1_1 = \cO_1 \,,      \nn\\
&&\hOm^{\pm}_1 = \cO_1 \,,\quad\, \hOm^1_1 = \cO_1  \,.    \lab{2.4}
\eea

After fixing the lowest order terms of $\hB^i_1$ and $\hOm^i_1$, we are
left with $18-6=12$ variables subject to the equations of motion. As we
already explained, only the lowest order terms of our variables $\hB$,
$\hOm$ can contribute to the boundary dynamics. This means that only
those field equations that contain zero modes of the remaining
variables are of interest to us. There are exactly 12 such equations.
Two of them stem from the constraints of the theory, and they read:
\bsubeq\lab{2.5}
\bea
&&\hOm^-_+ -\hOm^-_-=\cO_1\,,             \nn\\
&&2\hB^+_--\ell\,\hOm^+_-=2\hB^+_+ -\ell\,\hOm^+_+ +\cO_1\,.\lab{2.5a}
\eea
The remaining 10 needed equations fall into two categories: 8
equations become asymptotic constraints, as their time derivatives
vanish at spatial infinity,
\bea
&&\hB^1_{\pm}=\cO_1\,,\quad \hB^+_+=\cO_1\,,
  \quad  \hB^-_-=\cO_1\,,                    \nn\\
&&\hOm^1_{\pm}=\cO_1\,, \quad \hOm^+_+=\cO_2\,,
  \quad \hOm^-_+=\cO_2\,,                               \lab{2.5b}
\eea
while the remaining two are proper dynamical equations:
\be
\pd_+\hB^+_-=\cO_1 \,, \qquad \pd_-\hB^-_+=\cO_1 \,.    \lab{2.5c}
\ee
\esubeq

As we can see, only the variables $\hB^+_-$ and $\hB^-_+$ have
independent non-vanishing zero modes, which means that the
boundary dynamics is essentially described by equations \eq{2.5c}.
The corresponding 2-dimensional theory carries one field degree of
freedom. Note that the variables $\hat B^+_-$ and $\hat B^-_+$ are
invariant under the action of the proper gauge transformations
on the boundary (Appendix A). Therefore, the form of the equations
\eq{2.5c} is independent of the particular gauge used.

The physical meaning of the fields $\hB^-_+$ and $\hB^+_-$ is seen by
the comparison with the energy and angular momentum expressions,
defined as the $r\to\infty$ limit of the following boundary integrals
\cite{15}:
$$
E=\int_0^{2\p}\cE^1 d\vphi \,,\qquad
M=\int_0^{2\p}\cM^1 d\vphi \,,
$$
where
\bea
\cE^1&=&2a\left(\om^0{_2}
        +\frac{1}{\ell}b^2{_2}
        -\frac{1}{\ell}b^0{_2}\right)b^0{_0}\,,            \nn\\
\cM^1&=&-2a\left(\om^2{_2}
        +\frac{1}{\ell}b^0{_2}
        -\frac{1}{\ell}b^2{_2}\right)b^2{_2}\,.             \nn
\eea
Using the asymptotic field equations \eq{2.5}, one finds
\bea
&&\cE_+\equiv\frac{1}{2}\left(\ell\cE^1+\cM^1\right)
       =-2a\hB^-_+ +\cO_1 \, ,\nn\\
&&\cE_-\equiv\frac{1}{2}\left(\ell\cE^1-\cM^1\right)
       =-2a\hB^+_- +\cO_1 \, .                              \lab{2.6}
\eea
Thus,
\bitem
\item[] the variables $\hB^-_+$ and $\hB^+_-$ are proportional
to the boundary values of the components $\cE_+$ and $\cE_-$ of the
energy/angular momentum densities.
\eitem
Rewritten in terms of $\cE_\pm$, the boundary field equations \eq{2.5c}
read: $\pd_{\pm} \cE_\mp=\cO_1$. The densities $\cE_\pm$ are invariants
of the proper gauge group, but transform non-trivially under the
conformal transformations \cite{15}:
\bea
&&\d_0\cE_+=-T^+\pd_+\cE_+ -2\pd_+T^+\cE_+
               +a\ell\,\pd^3_+T^+ +\cO_1\, ,  \nn\\
&&\d_0\cE_-=-T^-\pd_-\cE_- -2\pd_-T^-\cE_-
               +a\ell\,\pd^3_-T^- +\cO_1\,,                 \lab{2.7}
\eea
where $T^{\pm}\equiv T\pm S$. These relations are equivalent to
equations (6.19) in Ref. \cite{15}, which tells us that the boundary
dynamics is described as the conformal field theory with classical
central charge $c=24\pi a\ell=3\ell/2G$. In section IV, we shall
explicitly establish the equivalence of our boundary dynamics with the
Liouville theory.

\section{Chern-Simons formulation of the teleparallel 3D gravity}

A particularly interesting feature of Riemannian 3D gravity is the fact
that it is equivalent to an ordinary gauge theory with a specific
interaction of the Chern-Simons type \cite{3}. This result has been
used in Ref. \cite{4} to show that the related boundary dynamics
coincides with the Liouville theory. In this section, we shall prove
that the teleparallel gravity can also be represented as a Chern-Simons
theory.

The canonical analysis of the teleparallel theory \eq{1.2} yields the
following expressions for its gauge symmetries \cite{15}:
\bea
&&\d_0 b^i{_\m}=\nabla_\m\ve^i
                -\frac{2}{\ell}\ve^i{}_{jk}b^j{_\m}\ve^k
                +\ve^i{}_{jk}b^j{_\m}\t^k\, ,               \nn\\
&&\d_0\om^i{_\m}=\nabla_\m\t^i\, .                          \lab{3.1}
\eea
Introducing the new parameters $\xi^\m$ and $\th^i$ by $\ve^i=-\xi^\m
b^i{_\m}$, $\t^i=-(\th^i+\xi^\m\om^i{_\m})$, one finds that these gauge
transformations are on-shell equivalent to the gauge transformations
\eq{1.1a}.  Witten's idea that the action of 3D gravity is a
combination of two pieces, each of which depends on an independent
gauge field, can be tested at the level of gauge transformations, where
we expect these two gauge fields to transform independently of each
other. After introducing the new variables \bsubeq\lab{3.2}
\be
A^i{_\m}=\om^i{_\m}\, ,\qquad
\bA^i{_\m}=\om^i{_\m}-\frac{2}{\ell}\,b^i{_\m}\, ,         \lab{3.2a}
\ee
the gauge transformations \eq{3.1} can be rewritten as
\bea
&&\d_0 A^i{_\m}=\pd_\m u^i+\ve^i{}_{jk}A^j{_\m}u^k\, ,
  \qquad u^i\equiv\t^i\, ,\nn\\
&&\d_0 \bA^i{_\m}=\pd_\m\bu^i+\ve^i{}_{jk}\bA^j{_\m}\bu^k \, ,
  \qquad \bu^i\equiv \t^i-\frac{2}{\ell}\,\ve^i\,.         \lab{3.2b}
\eea
\esubeq

The fields $A$ and $\bA$ are recognized as the gauge fields of two
independent $SL(2,R)$ gauge groups. Indeed, the fundamental matrix
representation of the $SL(2,R)$ generators (real, traceless,
$2\times 2$ matrices) can be chosen as $T_0=\frac{1}{2}i\s_2$,
$T_1=\frac{1}{2}\s_3$, $T_2=\frac{1}{2}\s_1$ ($\s_i$ are the Pauli
matrices), i.e.
$$
T_0=\frac{1}{2}\left(\ba{cc}
                     0 & 1 \\
                    -1 & 0 \\
                     \ea\right)\, ,\qquad
     T_2=\frac{1}{2}\left(\ba{cc}
                          0 & 1 \\
                          1 & 0 \\
                          \ea\right)\, ,\qquad
          T_1=\frac{1}{2}\left(\ba{cc}
                               1 & 0  \\
                               0 & -1 \\
                               \ea\right)\, .
$$
The Cartan metric has the form
$\g_{ij}=-2\Tr(T_iT_j)=\eta_{ij}=(+1,-1,-1)$, and the Lie algebra is
given as $[T_i,T_j]=\ve_{ij}{}^kT_k$, with $\ve_{012}=1$. Thus, the
original gauge symmetry \eq{3.1} can be equivalently represented by a
set of {\it two independent\/} $SL(2,R)$ gauge symmetries.

Using the variables \eq{3.2a}, we can rewrite the original
field equations \eq{1.3} in the form
\be
F^i{}_{\m\n}(A)=0\, ,\qquad F^i{}_{\m\n}(\bA)=0\,,         \lab{3.3}
\ee
where $F^i{}_{\m\n}(A)=\pd_\m A^i{_\n}-\pd_\n A^i{_\m}
+\ve^i{}_{jk}A^j{_\m}A^k{_\n}$ is the $SL(2,R)$ gauge field
strength.

The existence of two gauge fields with independent transformation laws
\eq{3.2b} suggests that the gravitational Lagrangian, when expressed in
terms of the new variables \eq{3.2a}, consists of two pieces, each of
which is a functions of only $A$ or $\bA$. To verify this idea, we
perform the change of variables \eq{3.2a} in the gravitational
Lagrangian \eq{1.2}, and obtain the important identity
\bsubeq\lab{3.4}
\be
\cL_G+\pd_\m D^\m=\cL_\cs(A)-\cL_\cs(\bA)\, ,              \lab{3.4a}
\ee
where $\cL_\cs$ is the Chern-Simons Lagrangian for $SL(2,R)$:
\bea
&&\cL_\cs(A)=\k\ve^{\r\m\n}\left(A^i{_\r}\pd_\m A_{i\n}
    +\frac{1}{3}\ve_{ijk}A^i{_\r}A^j{_\m}A^k{_\n}\right)\, ,\nn\\
&&D^\m=-\k\ve^{\m\n\r}\bA^i{_\n}A_{i\r}
      = a\ve^{\m\n\r}b^i{_\n}\om_{i\r}\, ,
         \quad \k\equiv\frac{a\ell}{2}\, .                \lab{3.4b}
\eea
\esubeq

Now, we wish to clarify the meaning of the total derivative term
appearing on the left hand side of \eq{3.4a}. Our asymptotic conditions
\eq{2.1} imply the following relations:
\be
A^i_+=\cO_1\, ,\qquad \bA^i_-=\cO_1\, .                       \lab{3.5}
\ee
It is well known that under such boundary conditions
both Chern-Simons actions $I_\cs(A)$ and $I_\cs(\bA)$ are
differentiable (see, e.g. Refs. \cite{12,25}). Then, as a
consequence of the identity \eq{3.4a}, we can define the improved
gravitational action,
\be
\tI_G=I_G +\D \, ,\qquad
\D\equiv a\int d^3x
     \pd_\m\left(\ve^{\m\n\r}b^i{_\n}\om_{i\r}\right)\, , \lab{3.6}
\ee
which is also differentiable. Thus, the role of the boundary term
$\D$ is to make the gravitational action differentiable under the
variations which respect the boundary conditions \eq{3.5}. The
identity \eq{3.4a} can now be written in the simple form
\be
\tI_G=I_\cs(A)-I_\cs(\bA)\, ,                              \lab{3.7}
\ee
showing that
\bitem
\item[] the improved gravitational action in the teleparallel 3D
gravity is equal to the difference of two Chern-Simons actions.
\eitem
Thus, the Chern-Simons formulation of 3D gravity is possible not only
in Riemannian GR \cite{4}, but also in the teleparallel gravity. The
spacetime with torsion is an equally acceptable arena for formulating
and discussing various features of boundary dynamics.

The result \eq{3.7} implies that the classical central charge of the
teleparallel theory has the value $c=12\cdot 4\pi\k={3\ell}/{2G}$,
in agreement with Ref. \cite{15}.

\section{Boundary dynamics 2}

In this section we study the role of boundary conditions on the form of
solutions of the teleparallel gravity in the Chern-Simons formulation,
and identify the related boundary dynamics as the Liouville theory.

With the adopted conventions for the $SL(2,R)$ generators, the gauge
potential $A^i\equiv A^i{_\m}dx^\m$ can be represented in the matrix
form as
$$
T_iA^i\equiv A=\frac{1}{2}\left(\ba{cc}
                           A^1 &  A^+ \\
                          -A^- & -A^1 \\
                          \ea\right) \, ,
$$
with the group indices written always as superscripts.

\subsection{Boundary conditions  for \mb{A^i{_\m}}}

\prg{1.} The asymptotic behavior of $b^i{_\m}$ and $\om^i{_\m}$ implies
the following conditions on the new variables \eq{3.2a}:
\be
A=\frac{1}{2}\left(\ba{cc}
    \dfrac{dr}{r}+\cO_2 &~~\cO_1                        \\[1.8ex]
   -\dfrac{2r}{\ell}dx^- +\cO_1 &~~-\dfrac{dr}{r}+\cO_2 \\[1.8ex]
             \ea\right) \, .                               \lab{4.1}
\ee
In addition to this, we also have some more precise details about
$A_1$:
\be
A^\pm_1=\cO_4\, ,\qquad
A^1_1=\frac{1}{r}+\cO_3\, .                                \lab{4.2}
\ee

\prg{2.} The field equations \eq{3.3} can be solved for $A$ as
\be
A=G^{-1}dG\, ,\qquad G\in SL(2,R)\, .                      \lab{4.3}
\ee
If the spatial sections of spacetime contain no holes, this solution is
valid also globally \cite{5}. Since we are interested only in the
asymptotic dynamics of gravity, we shall ignore the presence of holes.

Although $G$ is introduced as an arbitrary element of $SL(2,R)$,
boundary conditions produce non-trivial restrictions on its form. To
investigate these restrictions we note that, without loss of
generality, the matrix $G$ can be represented as
\be
G=g\r\, ,\qquad \r\equiv\left(\ba{cc}
                     \sqrt{r/\ell} & 0       \\[1.2ex]
                     0  & \sqrt{\ell/r}      \\[1.2ex]
                             \ea\right) \, ,               \lab{4.4}
\ee
where $g$ is also in $SL(2,R)$. Then, the relation
$$
A=G^{-1}dG=\r^{-1}\cA\r+\r^{-1}d\r\, ,\qquad \cA\equiv g^{-1}dg \, ,
$$
and the asymptotic conditions \eq{4.1} for $A$, define the asymptotic
behavior of $\cA$:
\be
\cA=g^{-1}dg=\left(\ba{cc}
                           0 & \cO_0 \\
                       -dx^- &~~0    \\
                 \ea\right)+\cO_2 \, .                      \lab{4.5}
\ee
Taking into account the additional information \eq{4.2} we find that
$\cA_1=\cO_3$, which implies that the $\cO_0$ term in $\cA$ does not
depend on $r$. Hence,
\bitem
\item[$a$)] to leading order in $1/r$, $\cA$ depends only on $t$ and
$\vphi$: $\cA=\cA(t,\vphi)$.
\eitem
In other words, $\cA$ is the field on the boundary. This result has a
direct consequence on the form of $g$ (Appendix B):
\bitem
\item[$b$)] to leading order in $1/r$, the matrix $g$  depends only
on $x^-$:
\be
g=g(x^-) \, .                                               \lab{4.6}
\ee
\eitem
As a consequence, the $\cO_0$ term in \eq{4.5} must be a function of
$x^-$, and we have
\bea
i) &&\qquad \cA_+=0\, ,    \nn\\
ii)&&\qquad\pd_+\cA^+_-=0\, ,
     \qquad \cA^-_-= 2\, ,
     \qquad \cA^1_-=0\, .                                  \lab{4.7}
\eea
The higher order terms are not written explicitly, but are understood.
Also, the term $\cA_1$ has been dropped, as it trivially vanishes on
the account of \eq{4.6}.

Thus, after the asymptotic conditions are taken into account, only one
element of the original matrix $G$ in \eq{4.3} is left to describe
the boundary dynamics:  $\cA^+_-=\cA^+_-(x^-)$.

\subsection{Boundary conditions for \mb{\bA^i{_\m}}}

\prg{1.} Applying the same procedure to the variable $\bA$, we obtain
\be
\bA=\frac{1}{2}\left(\ba{cc}
    -\dfrac{dr}{r}+\cO_2 &~~-\dfrac{2r}{\ell}dx^+ +\cO_1 \\[1.8ex]
            \cO_1 &~~\dfrac{dr}{r}+\cO_2                 \\[1.8ex]
             \ea\right) \, ,                               \lab{4.8}
\ee
and also
\be
\bA^\pm_1=\cO_4\, ,\qquad
\bA^1_1=-\frac{1}{r}+\cO_3\, .                             \lab{4.9}
\ee

\prg{2.} The general solution for $\bA$ is given as
\be
\bA=\bG^{-1}d\bG\, ,\qquad \bG\in SL(2,R)\, .              \lab{4.10}
\ee
In order to clarify the effect of boundary conditions on the form of
$\bG$, we write
\be
\bG=\bg \r^{-1}\, .                                        \lab{4.11}
\ee
Then, the relation
$$
\bA=\bG^{-1}d\bG=\r\bcA\r^{-1}+\r d\r^{-1}\, ,
    \qquad \bcA=\bg^{-1}d\bg\, ,
$$
and the asymptotic form \eq{4.8} of $\bA$, lead to
\be
\bcA=     \left(\ba{cc}
               0  & -dx^+ \\
               \cO_0 &~~0 \\
               \ea\right) +\cO_2\, .                      \lab{4.12}
\ee
The additional conditions \eq{4.9} imply that $\bcA_1=\cO_3$; hence,
\bitem
\item[$a$)] to leading order in $1/r$, $\bcA$ depends only on $t$ and
$\vphi$:~~$\bcA=\bcA(t,\vphi)$.
\eitem
This result, in return, yields a precise information on the matrix
$\bg$ itself (Appendix B):
\bitem
\item[$b$)] to leading order in $1/r$, the matrix $\bg$ depends only
on $x^+$:
\be
\bg=\bg(x^+) \, ,                                          \lab{4.13}
\ee
\eitem
Consequently, we have
\bea
i)&&\qquad \bcA_-=0\, ,                                  \nn\\
ii)&&\qquad\pd_-\bcA^-_+=0\, ,
     \qquad\bcA^+_+= -2 \, ,
     \qquad\bcA^1_+=0\, .                                 \lab{4.14}
\eea
The only element of the matrix $\bG$ in \eq{4.10} that corresponds to a
non-trivial boundary dynamics is $\bcA^-_+=\bcA^-_+(x^+)$.

We see that the whole content of our theory reduces to a
2-dimensional field theory at the boundary. The latter is
described by the $SL(2,R)$ matrices $g(x^-)$ and $\bg(x^+)$,
subject to the conditions \eq{4.7} and \eq{4.14}. In what follows,
this 2-dimensional theory will be analyzed in details.

\subsection{Liouville dynamics at the boundary}

The result of the previous analysis is completely expressed by Eqs.
\eq{4.7} and \eq{4.14}: after imposing the asymptotic conditions,
there are only two components of the fields $A$ and $\bA$ that remain
non-trivial at the boundary: $\cA^+_-(x^-)$ and $\bcA^-_+(x^+)$.

With a convenient parametrization, these {\it two independent chiral
fields\/} can be associated to a {\it single Liouville field\/}
$\phi=\phi(x^-,x^+)$. To show that, we introduce the new matrix
$\g=g^{-1}\bg$, and the related ``currents" $K=\g^{-1}d\g$
and $\bK=\g d\g^{-1}$. Then we find
\bea
&&K_+=\g^{-1}\pd_+\g=\bg^{-1}\pd_+\bg=\bcA_+\, , \nn\\
&&\bK_-=\g\pd_-\g^{-1}=g^{-1}\pd_- g=\cA_-\, ,             \lab{4.15}
\eea
and the Eqs. \eq{4.7} and \eq{4.14} lead to
\bsubeq\lab{4.16}
\bea
&&\pd_- K_+=0\, , \qquad \pd_+\bK_-=0\, ,              \lab{4.16a}\\
&&K^+_+= -2 \, ,\qquad\, \bK^-_-= 2 \, ,               \lab{4.16b}\\
&&K^1_+=0\, ,\qquad\quad \bK^1_-=0\, .                 \lab{4.16c}
\eea
\esubeq
In return, the conditions \eq{4.16} imply \eq{4.7} and \eq{4.14} up
to a constant $SL(2,R)$ matrix. This means that the complete content
of the equations \eq{4.7} and \eq{4.14} is basically expressed in
terms of $K$ and $\bK$.

For arbitrary $\g$, the boundary conditions $\pd_- K_+=0$ represent
the set of three field equations for the Wess-Zumino-Novikov-Witten
(WZNW) model; they are equivalent to $\pd_+\bK_-=0$. After
introducing the Gauss coordinates $(X,\phi,Y)$ to parametrize the
$SL(2,R)$ group manifold, these equations, combined with \eq{4.16b},
reduce to the Liouville equation (Appendix C),
\bsubeq
\be
\pd_-\pd_+\phi-2e^\phi=0\, ,                            \lab{4.17a}\\
\ee
while the remaining two conditions \eq{4.16c} yield:
\be
2X=-\pd_-\phi\, ,\qquad 2Y=\pd_+\phi\, .                \lab{4.17b}
\ee
\esubeq

The general solution for the Liouville field is given in terms of {\it
two chiral functions\/}, hence we have the correct balance of the
dynamical degrees of freedom at the boundary. Therefore,
\bitem
\item[] the boundary dynamics of the teleparallel gravity
can be identified with Liouville theory.
\eitem

\subsection{Energy-momentum tensor}

The non-trivial fields at the boundary, $\bcA^-_+$ and $\cA^+_-$, can
be expressed in terms of the Liouville field as follows:
\bea
&&\bcA^-_+=K^-_+ =
        \frac{1}{2}(\pd_+\phi)^2-\pd_+^2\phi=\Th_{++}\, , \nn\\
&&\cA^+_-=\bK^+_- = -\frac{1}{2}(\pd_-\phi)^2+\pd_-^2\phi
                  =-\Th_{--}\, ,   \nn
\eea
where $\Th_{\mp\mp}$ are components of the Liouville energy-momentum
tensor. Then, we can return to the original gravitational variables
discussed in section II,
$$
\cA^+_-=\hOm^+_- =\frac{2}{\ell}\,\hat B^+_- \, ,\qquad
\bcA^-_+=\hOm^-_+ -\frac{2}{\ell}\,\hat B^-_+
        =-\frac{2}{\ell}\,\hat B^-_+ \, ,
$$
and derive the relations
\be
\Th_{++}=-\frac{2}{\ell}\,\hB^-_+\, ,\qquad
\Th_{--}=-\frac{2}{\ell}\,\hB^+_-\, .                    \lab{4.18}
\ee
Using the known symmetry of the Liouville theory,
$\d_0\phi=-\ve\cdot\pd\phi-\pd\cdot\ve$, with $\ve=(T^+,T^-)$, one
finds the transformation law
\bea
&&\d_0\Th_{++}=-T^+\pd_+\Th_{++}
               -2\pd_+T^+\Th_{++} +\pd_+^3T^+\, , \nn\\
&&\d_0\Th_{--}=-T^-\pd_-\Th_{--}
               -2\pd_-T^-\Th_{--} +\pd_-^3T^-\, , \nn
\eea
in agreement with \eq{2.7}. Thus,
\bitem
\item[] the boundary fields $\hB^-_+(x^+)$ and $\hB^+_-(x^-)$
are proportional to the $(++)$ and $(--)$ components of
the Liouville energy-momentum tensor.
\eitem
This agrees with the result of Ref. \cite{8}, which is, however, based
on purely geometric arguments.

\section{Concluding remarks}

In the present paper we analyzed the nature of asymptotic dynamics in
the teleparallel 3D gravity, starting from the asymptotic AdS
configuration \eq{2.1} for the triad $b^i{_\m}$ and the connection
$\om^i{_\m}$. The full set of gauge symmetries is naturally split into
two parts: the conformal symmetry and the proper gauge symmetry
\cite{15}.

Using the asymptotic field equations, we found that there are only two
non-trivial dynamical modes at the boundary: $\hB^+_-(x^-)$ and
$\hB^-_+(x^+)$. Their transformation laws show that the boundary
dynamics is described as a conformal field theory with the classical
central charge $c=3\ell/2G$.

An important step in clarifying the structure of the boundary
field theory is made by showing that the improved action of the
teleparallel 3D gravity is given as a difference of two
Chern-Simons actions. This result leads us to recognize Liouville
theory as the dynamical structure at the boundary. The fact that
the same result is found also in Riemannian 3D gravity
\cite{4,5,6,7,8,9,10,11} indicates very strongly that the nature
of boundary dynamics  does not depend on the {\it geometric
context\/}, but only on the {\it boundary conditions\/}.

\acknowledgements
One of us (M.B.) would like to thank Friedrich Hehl for reading
the manuscript. This work was partially supported by the Serbian
Science foundation, Serbia.

\appendix

\section{Proper gauge transformations}

Using the notation $\hat\d \equiv \d_0 (T=S=0)$ for the proper gauge
variation, the general transformation law \eq{1.1a} leads to the
following asymptotic relations:
$$
\ba{ll}
\hat\d\hB^+_-=\hat\d\hB^-_+ = 0\,,             &
\hat\d\hOm^+_-=\hat\d\hOm^-_+ = 0\,,                     \\[1.4ex]
\hat\d\hB^+_+=\ell\hat\th-2\hat\xi^1\,,        &
\hat\d\hOm^+_+ = 0 \,,                                   \\[1.4ex]
\hat\d\hB^-_-=-\ell\hat\th\,,                  &
\hat\d\hOm^-_-=-2\hat\th\,,                              \\[1.4ex]
\hat\d\hB^1_+=-\pd_+\hat\xi^1
       +\dfrac{\ell}{2}\hat\th^-\,, \quad      &
\hat\d\hOm^1_+=-\pd_+\hat\th\,,                          \\[1.4ex]
\hat\d\hB^1_-=-\pd_-\hat\xi^1
              -\dfrac{\ell}{2}\hat\th^+ \,,    &
\hat\d\hOm^1_-=-\pd_-\hat\th-\hat\th^+\,,                \\[1.4ex]
\hat\d\hB^+_1=4\hat\xi^+-\hat\th^+\,,          &
\hat\d\hOm^+_1 = \dfrac{2}{\ell}\hat\th^+\,,             \\[1.4ex]
\hat\d\hB^-_1 = 4\hat\xi^- + \hat\th^- \,,     &
\hat\d\hOm^-_1 =\dfrac{4}{\ell}
        \left(\hat\th^-+2\hat\xi^-\right)\,,             \\[1.4ex]
\hat\d\hat B^1_1=\dfrac{2}{\ell}\hat\xi^1\,,   &
\hat\d\hat\Om^1_1=\dfrac{2}{\ell}\hat\th \,,
\ea
$$
where $\,\hat\th \equiv \hat\th^1+\hat\xi^1/\ell$. The equality in
the above expressions is the equality up to $\cO_1$ terms (which
are omitted for simplicity). The last six equations show that the
proper gauge freedom can be fixed as in \eq{2.4}.

\section{Conditions \eq{4.6} and \eq{4.13}}

The general expression \eq{4.5} for $\cA=\cA(t,\vphi)$ can be written
as follows:
\bea
&&\cA_+=g^{-1}\pd_+ g
        =\left(\ba{cc}
               0 & \pd_+\a \\
               0 & 0 \\
               \ea\right) \, ,             \nn\\
&&\cA_-=g^{-1}\pd_- g
        =\left(\ba{cc}
                0 & \b \\
               -1 & 0  \\
               \ea\right) \, ,             \nn
\eea
where $\a$ and $\b$ are functions of $t$ and $\vphi$. The first
condition implies
$$
g=U \left(\ba{cc}
          1 & \a \\
          0 & 1  \\
          \ea\right) \, ,\qquad U=U(x^-)\, .
$$
Substituting this expression into the second condition, one finds
$$
U^{-1}\pd_-U=\left(\ba{cc}
                   -\a &~~\b+\a^2-\pd_-\a \\
                   -1  &~~\a              \\
                   \ea\right) \, .
$$
Consequently, $\a=\a(x^-)$, $\b=\b(x^-)$, which proves \eq{4.6}.

Similarly, the expression \eq{4.12} for $\bcA=\bcA(t,\vphi)$
yields
\bea
&&\bcA_-=\bg^{-1}\pd_-\bg
        =\left(\ba{cc}
                         0 & 0 \\
               \pd_-\bar\a & 0 \\
               \ea\right) \, ,              \nn\\
&&\bcA_+=\bg^{-1}\pd_+\bg
        =\left(\ba{cc}
                0 & -1     \\
                \bar\b & 0 \\
               \ea\right) \, ,              \nn
\eea
with $\bar\a=\bar\a(t,\vphi)$, $\bar\b=\bar\b(t,\vphi)$. The first
condition implies
$$
\bg=\bar U \left(\ba{cc}
                  1 & 0  \\
                 \a & 1  \\
                 \ea\right) \, ,\qquad \bar U=\bar U(x^+)\, .
$$
Then, the substitution of this expression into the second condition
leads to $\bar\a=\bar\a(x^+)$, $\bar\b=\bar\b(x^+)$, which proves
\eq{4.13}.

\section{From WZNW to Liouville}

In this appendix we recall the process of reduction of the WZNW field
equations for the group $SL(2,R)$, to Liouville equation.

Every element $\g$ of $SL(2,R)$, in a neighborhood of identity, admits
the Gauss decomposition:
$$
\g=\left(\ba{cc}
         1 & X \\
         0 & 1 \\
         \ea\right) \left(\ba{cc}
                          e^{\phi/2} & 0 \\
                          0 & e^{-\phi/2}\\
                          \ea\right) \left(\ba{cc}
                                           1 & 0 \\
                                           Y & 1 \\
                                           \ea\right) \, ,
$$
where  $(X,\phi,Y)$ are group coordinates. With the above expression
for $\g$, the Lie algebra valued 1-form $K=\g^{-1}d\g=T_iK^i$, defines the
quantity $K^i$, the vielbein on the group manifold:
\bea
&&K^+=2e^{-\phi}dX \, ,\nn \\
&&K^1=2Ye^{-\phi}dX +d\phi \, ,\nn \\
&&K^-=-2\left(-Y^2e^{-\phi}dX -Yd\phi +dY\right) \, . \nn
\eea
Similarly, $\bK=\g d\g^{-1}=T_i\bK^i$ defines $\bK^i$:
\bea
&&\bK^+=2\left(-dX+Xd\phi+X^2e^{-\phi}dY\right)  \, ,\nn \\
&&\bK^1= -d\phi-2Xe^{-\phi}dY \, ,\nn \\
&&\bK^-=2e^{-\phi}dY \, . \nn
\eea
The relations $\pd_-K^i_+=0$ are equivalent to $\pd_+\bK^i_-=0$,
and produce the field equations of the WZNW model:
\bea
&&\pd_-\left(e^{-\phi}\pd_+X\right)=0\, ,      \nn\\
&&\pd_-\pd_+\phi+2\pd_-Y\pd_+Xe^{-\phi}=0\, ,  \nn\\
&&\pd_+\left(e^{-\phi}\pd_-Y\right)=0\, .      \nn
\eea
Under the boundary conditions \eq{4.16b}, the second of these equations
is easily seen to reduce to the Liouville equation \eq{4.17a}.

\end{document}